\documentclass[aps,prb,twocolumn,showpacs,floatfix,superscriptaddress]{revtex4}
\usepackage{graphicx,bm,amsmath,amssymb,natbib,url,epsfig,psfrag}

\begin{document}

\title{Fano resonance in a two-level quantum dot
side-coupled to leads}

\author{W.-R. Lee}
\author{Jaeuk U. Kim}
\author{H.-S. Sim}
\affiliation{Department of Physics, Korea Advanced Institute of
Science and Technology, Daejeon 305-701, Korea}

\date{\today}

\begin{abstract}
We theoretically study
Fano resonance in a two-level quantum dot
side-coupled to two leads, which are connected by a direct channel.
% by using Keldysh formalism and a self-consistent Hartree-Fock approach.
The resonance lineshape is found to be deformed,
from the conventional Fano form,
%from the well-known single-level lineshape,
by interlevel Coulomb interaction and interlevel interference.
We derive the connection between the lineshape deformation
and the interaction-induced nonmonotonicity of level occupation,
which may be useful for experimental study.
% of level occupation.
The dependence of the lineshape on the transmission
of the direct channel and
on the dot-lead coupling matrix elements is discussed.

\end{abstract}

%72.10.-d Theory of electronic transport; scattering mechanisms
%73.63.Kv Quantum dots
%73.23.Hk Coulomb blockade

\maketitle

%\section{Introduction}

Fano resonance, \cite{Fano61} which
is the interference between a resonant state and a continuum,
appears ubiquitously in various systems.
It has a lineshape of the form
\begin{eqnarray}\label{FanoLineshape}
T (\varepsilon,q) \sim {|\varepsilon+q|^2\over\varepsilon^2+1}.
\end{eqnarray}
Here, $\varepsilon= (\omega - \xi_0)/\Gamma$ is the detuning
parameter measuring energy $\omega$ from the resonance center $\xi_0$ and
normalized by the resonance half-width $\Gamma$, and $q$ is the Fano
parameter characterizing lineshape asymmetry. For
$q\rightarrow\infty$ Eq.~\eqref{FanoLineshape} becomes the
Breit-Wigner form, while for $q = 0$ it shows an antiresonance. In
general, $q$ is a complex quantity.~\cite{Clerk01}

Recently, Fano resonance has been investigated
%theoretically and experimentally
in mesoscopic electron systems such as
waveguides,~\cite{Tekman93} quantum
dots,~\cite{Gores00,Zacharia01} Aharonov-Bohm rings coupled to a
quantum dot,~\cite{Kobayashi02,Kobayashi04,Fuhrer06} and carbon
nanotubes.~\cite{Kim03,Yi03} The studies imply that Fano
resonance provides a useful tool studying
dephasing.~\cite{Clerk01} On the other hand, some aspects of the
interplay between Fano resonance and Coulomb interaction have been
studied. They include Fano resonance modified by charge
sensing~\cite{Johnson04,Stefanski05} and Fano-Kondo antiresonance in
a spin-degenerate single-level quantum
dot.~\cite{Bulka01,Hofstetter01,Sato05}

The Fano lineshape~\eqref{FanoLineshape} is applicable for a system
with a {\em single} resonant level. It is valid as well for
multilevel systems as long as each single-particle
level is well separated
from the adjacent levels in energy. Most studies on Fano resonance
%, including the above-mentioned examples,
have been carried out mainly in this single-level regime. However,
one may often find the multilevel regime where single-particle level
spacing is comparable to level broadening. In this regime, the
single-level Fano form is not applicable any more. Moreover, this
regime possesses interesting effects, absent in the single-level
regime, such as nonmonotonic level
occupation~\cite{Berkovits05,Konig05,Sindel05}
%(or level occupation inversion)
%in which the occupation of a single-particle level can be smaller
%than that of the adjacent upper levels
due to Coulomb repulsion. It has been reported~\cite{Goldstein06}
that the nonmonotonic behavior of level occupation influences
Breit-Wigner lineshape. Therefore, it may be interesting to see the
modification of resonance lineshape, from Eq.~\eqref{FanoLineshape},
in a more general {\em multilevel Fano} regime and to analyze the
influence of Coulomb interaction on resonance lineshape,
which is the aim of the present paper.

\begin{figure}[b]
\includegraphics[width=0.40\textwidth]{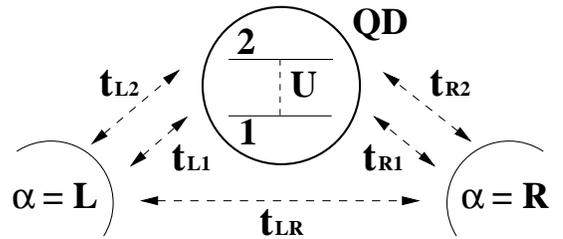}
\caption{A quantum dot with spinless two levels $\gamma = 1,2$,
side-coupled, with coupling matrix element $t_{L(R) \gamma}$, to two
leads $L$ and $R$, which are connected by a direct channel with
coupling $t_{LR}$. There is Coulomb repulsion $U$ between the two
levels.} \label{1Setup}
\end{figure}

In this work, we theoretically study Fano resonance in a two-level
electron quantum dot (QD) side-coupled to two leads, which are
connected by a direct channel (see Fig.~\ref{1Setup}). Interlevel
Coulomb repulsion in the dot is taken into account and the spin of
electrons is neglected for simplicity. We use Keldysh formalism and
a self-consistent Hartree-Fock (SCHF) approach, to obtain and to
analyze the Fano resonance lineshape of the two-level system.
%in the linear response regime.
% near zero temperature.
The two-level lineshape is found to be deformed from the
single-level form~\eqref{FanoLineshape} by the Coulomb repulsion and
interlevel interference. We derive the connection
[Eqs.~\eqref{connection1} to~\eqref{connection2}] between the
lineshape deformation and the nonmonotonicity of level occupation,
when the QD level spacing (after renormalized by the repulsion) is
larger than level broadening so that the nonmonotonicity is not too
strong. The connection may be useful for experimental study.
%on the nonmonotonic level occupation.
We also discuss the dependence of the nonmonotonicity on the
direct-channel coupling, an extension to a spinful single-level
case, and the temperature range where the SCHF result is valid,
below which correlation-induced
resonances~\cite{Meden06,Karrasch06,Lee07,Kashcheyevs07,Silvestrov07}
may emerge.

%\section{Model}

We start with the Hamiltonian $H_D$ of the spinless two-level QD,
%The Hamiltonian of the spinless two-level QD is
\begin{eqnarray}
\nonumber \\
H_D & = & \sum_{\gamma=1,2}(\xi_{\gamma}-eV_g)
d_{\gamma}^{\dag}d_{\gamma} +
U d^\dagger_1 d^\dagger_2 d_2 d_1,
%n_1n_2,
\label{QDHamiltonian}
\end{eqnarray}
where $d^\dagger_\gamma$ creates an electron at QD level $\gamma =
1,2$ with energy $\xi_\gamma$, $V_g$ is the gate voltage applied to
the QD, and $U$ is the interlevel Coulomb repulsion. Without loss of
generality, we can set $\xi_1 < \xi_2$. The QD couples via tunneling
to noninteracting leads $\alpha = L,R$, which are connected by a
direct channel (see Fig.~\ref{1Setup}), so that the total
Hamiltonian of the system is $H = H_D + H_L + H_{TD} + H_{TL}$,
where the
two leads, the dot-lead tunneling, and the lead-lead tunneling are
described by $H_L=\sum_{k,\alpha}\xi_{k\alpha}
c_{k\alpha}^{\dag}c_{k\alpha}$,
$H_{TD}=\sum_{k\alpha\gamma}(t_{\alpha\gamma}
c_{k\alpha}^{\dag}d_{\gamma}+\mathrm{h.c.})$, and
$H_{TL}=\sum_{kk'}(t_{LR}c_{kL}^{\dag}c_{k'R}+\mathrm{h.c.})$,
respectively. Here $c_{k\alpha}^\dagger$ creates an electron with
momentum $k$ and energy $\xi_{k \alpha}$ at lead $\alpha$.
We ignore
for simplicity the momentum dependence of tunneling matrix elements
$t_{\alpha \gamma}$ between level $\gamma$ and lead $\alpha$, and
that of $t_{LR}$ of the direct channel.

It is worthwhile to note that electron transport through the two
levels depends on the dot-lead coupling $t_{\alpha \gamma}$. To see
this, we consider simple cases~\cite{Silva02}
with time reversal symmetry where $t_{\alpha \gamma}$'s are chosen to be
real, $t_{L1} = t_{R1} = t_1 > 0$, $t_{L2} = s t_{R2} = t_2
> 0$, and the phase parameter
$s = \pm 1$.
%$s \equiv \mathrm{sgn}(t_{L1} t_{L2} t_{R1} t_{R2}) = \pm 1$.
After redefining a pair of two orthogonal leads,
saying $\tilde{\alpha} = \pm 1$,
%saying $\tilde{\alpha} = + 1$ and $\tilde{\alpha} = -1$,
by $\tilde{c}_{k \tilde{\alpha}}=(c_{kL} + \tilde{\alpha}
c_{kR})/\sqrt{2}$, one finds that
for $s = +1$, the two QD levels couple to the
same lead $\tilde{\alpha} = +1$, while for $s=-1$, they
couple to different leads each other.
%($\gamma = 1$ to $\tilde{\alpha}=+1$
%while $\gamma = 2$ to $\tilde{\alpha} = -1$).
This $s$-dependent nature of coupling to leads $\tilde{\alpha}$
characterizes~\cite{Silva02,Sindel05,Goldstein06}
electron transport such as interference.
Below,
we will use the above choice of $t_{\alpha \gamma}$'s
and see the dependence of the two-level Fano lineshape on $s$.
Note that $t_{LR}$ is also chosen to be real as well.

We obtain electric current of the QD system
using Keldysh formalism.~\cite{Meir92}
The current, $J_L = -e \langle \dot{n}_L \rangle
=(ie/\hbar) \langle [n_L, H ] \rangle$,
in the lead $L$ can be expressed as
\begin{equation}
\label{LeftCurrent}
J_L = \frac{2e}{\hbar}
\mathfrak{Re}
\int \frac{d \omega}{2 \pi}
\sum_{k\gamma} t_{L\gamma} G_{\gamma,kL}^<(\omega) +
\sum_{kk'} t_{LR} G_{k'R,kL}^<(\omega).
\end{equation}
Here, $n_L\equiv\sum_kc_{kL}^{\dag}c_{kL}$ is
the electron density operator,
and $G^< (\omega)$'s are lesser Green's functions
which correspond in time domain to
$G_{\gamma,kL}^< (t,t') \equiv i \langle c_{kL}^\dag(t') d_\gamma(t)\rangle$
and $G_{k'R,kL}^< (t,t')=i \langle c_{kL}^\dag(t') c_{k'R}(t) \rangle$.
One finds the current $J_R$ in lead $R$ in the same way.
After some algebra~\cite{Meir92} using
the relations connecting $G^<$'s and the retarded Green function
$G^r_{\gamma \gamma'}$ of the QD
[see Eqs. (\ref{GDiagonal}) and (\ref{GOffdiagonal})],
the wide-band approximation for
lead Green's function $g_{k \alpha}$,
$\sum_{k}g_{k\alpha}^r(\omega)\simeq-i\pi\rho$,
$\rho$ being the density of states of leads,
and the steady-state current conservation $J_L = - J_R$,
we arrive at a useful form of the current,
\begin{eqnarray}\label{Current}
J={e\over h}\int
d\omega\big(f_{L}(\omega)-f_{R}(\omega)\big)(\mathcal{T}_B
+\mathcal{T}_{QD}(\omega)),
\end{eqnarray}
where $f_\alpha$ is the Fermi distribution of lead $\alpha$.
The background transmission,
$\mathcal{T}_B = 4 \pi^2 \rho^2 t_{LR}^2 / x^2$,
comes only from the direct channel,
%The first term shows
%the background contribution only from the direct channel, $\mathcal{T}_B = 4
%\pi^2 \rho^2 t_{LR}^2 / x^2$,
where $x =1 + \pi^2 \rho^2 t_{LR}^2$
is the factor counting multiple reflections via the direct
channel.
The term $\mathcal{T}_{QD}$ results from the paths
passing through the QD, and it depends on the phase parameter $s$,
\begin{equation}
\mathcal{T}_{QD}^{s=+1} (\omega) = - \mathfrak{Im} \big[ y
\tilde{\Gamma}_{1} G_{11}^{r} + y \tilde{\Gamma}_{2} G_{22}^{r} + 2
y \sqrt{\tilde{\Gamma}_{1} \tilde{\Gamma}_{2}}G_{12}^{r} \big],
\label{QDTransmission_p}
\end{equation}
\begin{equation}
\mathcal{T}_{QD}^{s=-1}(\omega) = - \mathfrak{Im} \big[ y
\tilde{\Gamma}_{1} G_{11}^r + y^* \tilde{\Gamma}_{2} G_{22}^r + i 2
y \tilde{\Gamma}_{1} \tilde{\Gamma}_{2} G_{11}^r G_{22}^{r*} \big],
\label{QDTransmission_n}
\end{equation}
where
$y=(1 - i \pi \rho t_{LR})^4 / x^2$ is a
complex factor coming from the effect of the direct channel
and
\begin{equation}
\tilde{\Gamma}_\gamma = \frac{\Gamma_\gamma}{x}
= \frac{2 \pi \rho t^2_\gamma }{1 + \pi^2 \rho^2 t^2_{LR}}
\label{Broadening}
\end{equation}
is the QD level broadening.~\cite{Bulka01,Hofstetter01}
Notice that $\tilde{\Gamma}_\gamma$
becomes narrower, as $t_{LR}$ increases,
from the level broadening
$\Gamma_\gamma = 2 \pi \rho t^2_\gamma$ of the QD without the direct
channel.
The first two terms of Eqs.
(\ref{QDTransmission_p},\ref{QDTransmission_n})
describe the direct contribution through the QD level $\gamma$ as well as
the interference between the paths through the level $\gamma$ and the
direct channel, while the third
shows the interlevel interference between the paths through the two levels.

The derivation of $\mathcal{T}_{QD}$ depends on $s$ due to the
coupling nature to the leads $\tilde{\alpha}$.
For $s=-1$, $G_{12}^r$ does not appear in $\mathcal{T}_{QD}$,
and it is necessary to use Keldysh equation
$G^< = G^r \Sigma^< G^a$ for the dot lesser function.
We use the noninteracting form of the lesser self-energy $\Sigma^<$
of the dot coming from the lead-dot coupling,
therefore $\mathcal{T}_{QD}^{s=-1}$
in Eq.~\eqref{QDTransmission_n} is an approximate form valid
within the SCHF approach used below.
%near zero temperature and in the linear response regime.
On the other hand, for $s=+1$,
the Keldysh equation is not necessary in the derivation, thus
$\mathcal{T}_{QD}^{s=+1}$ in Eq.~\eqref{QDTransmission_p} is exact.
Note that Eqs.
(\ref{Current},\ref{QDTransmission_p},\ref{QDTransmission_n}) are
reduced into the forms found in the previous works on the
single-level Fano resonance~\cite{Bulka01,Hofstetter01}
(when $\xi_2 - \xi_1 \gg \tilde{\Gamma}_{\gamma=1,2}$) and on a
multilevel QD without the direct channel~\cite{Meir92} ($t_{LR} \to 0$).

We obtain the retarded QD Green's function $G^r_{\gamma \gamma'}$
%in Eqs.  (\ref{QDTransmission_p},\ref{QDTransmission_n})
and the level occupation $\langle n_{\gamma \gamma'} \rangle$
in equilibrium by using the equation of motion method
and the SCHF approach,
\begin{eqnarray}
G_{\gamma\gamma}^r (\omega) & = & \frac{1}{\mathbf{D}} (\omega -
\xi_{\bar{\gamma}} + eV_g - U \langle n_{\gamma} \rangle -
\Sigma_{\bar{\gamma}}^r),
\label{GDiagonal} \\
G_{12}^r (\omega) & = & \frac{1}{\mathbf{D}}
(-U \langle n_{12} \rangle + \Sigma_{12}^r),
\label{GOffdiagonal} \\
\langle n_{\gamma \gamma'} \rangle & = & - \frac{1}{\pi}
\int d \omega f(\omega) \mathfrak{Im} G_{\gamma' \gamma}^r
(\omega),
\label{Occupation}
\end{eqnarray}
where $\mathbf{D}=\big(\omega-\xi_1+eV_g-U\langle
n_2\rangle-\Sigma_1^r\big)\big(\omega-\xi_2+eV_g-U\langle
n_1\rangle-\Sigma_2^r\big)-\big(U\langle
n_{12}\rangle-\Sigma_{12}^r\big)^2$, $\bar{\gamma}$ means the
level different from $\gamma$, $\langle n_\gamma \rangle \equiv
\langle n_{\gamma \gamma} \rangle$ is the occupation of level
$\gamma$, and $f(\omega)=1/(e^{\beta\omega}+1)$. The self-energies
are found to be $\Sigma_{\gamma}^r=-(s^{\gamma-1} \pi \rho t_{LR} +
i)\tilde{\Gamma}_\gamma$ and $\Sigma_{12}^r=-\delta_{s,1}(\pi \rho
t_{LR} + i) \sqrt{\tilde{\Gamma}_{1}\tilde{\Gamma}_{2}}$;
from $\mathfrak{Im} \Sigma_\gamma^r$,
one can get Eq. (\ref{Broadening}).
We remark that
$\Sigma_{12}^r$ and $\langle n_{12} \rangle$,
therefore $G_{12}^r$, vanish for $s = -1$, because of
the $s$-dependent coupling nature to the leads $\tilde{\alpha}$.
The SCHF approach is
good when
%the level broadening
$\Gamma_\gamma$ is not too
large compared with level spacing $\xi_2 - \xi_1$.~\cite{Sindel05}
We later discuss the temperature
range where the SCHF result may be valid.
Since we have interest in resonance lineshape, we will focus on the
the linear response regime below.

\begin{figure}[htb]
\includegraphics[width=0.4\textwidth]{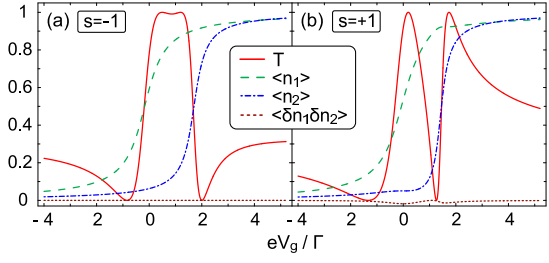}
\caption{(color online)
Two-level Fano
lineshape $\mathcal{T}$, level occupation $\langle n_\gamma
\rangle$, and occupation cross-correlation
$\langle \delta n_1 \delta n_2 \rangle$
as a function of $V_g$
for (a) $s=-1$ and (b) $s=+1$
in the noninteracting case of $U=0$.
We choose
$\Gamma_1/\Gamma=0.63$, $\Gamma_2/\Gamma=0.37$,
$(\xi_2-\xi_1)/\Gamma=1.6$, and $\pi\rho t_{LR}=0.3$,
where
%$\Gamma_\gamma = 2 \pi \rho t_\gamma^2$ and
$\Gamma\equiv\Gamma_1+\Gamma_2$.
Zero-temperature results of $\langle n_{\gamma \gamma'} \rangle$
and $\mathcal{T}$ are used for simplicity.
}\label{2NonintQD}
\end{figure}

%\section{Noninteracting cases}

We first discuss two-level resonance lineshape in the noninteracting
case of $U=0$. The two-level lineshape
can be obtained as $\mathcal{T} \equiv
\mathcal{T}_B + \mathcal{T}_{QD} (\omega = 0)$,
\begin{equation}\label{NoninteractingFano}
\mathcal{T} = \mathcal{T}_B \frac{ [ \varepsilon_1 \varepsilon_2
- s - (\zeta^2 -1) \delta_{s,1}
+ s q (\varepsilon_1 + s
\varepsilon_2 - 2 \zeta \delta_{s,1} ) ]^2}{(
\varepsilon_1 \varepsilon_2 - s - (\zeta^2 - 1) \delta_{s,1}
)^2 + (\varepsilon_1 + s \varepsilon_2 - 2 \zeta
\delta_{s,1})^2},
\end{equation}
where $\varepsilon_\gamma = (eV_g - \xi_\gamma) /
\tilde{\Gamma}_\gamma + s^{\gamma - 1} \pi \rho t_{LR}$ is the
detuning parameter of level $\gamma = 1,2$,
$\zeta \equiv -
(\tilde{\Gamma}_1 \tilde{\Gamma}_2)^{-1/2}
\mathfrak{Re}\Sigma^r_{12}
= \pi \rho t_{LR}\delta_{s,1}$,
the terms with Kronecker delta $\delta_{s,1}$ come from
$\Sigma_{12}^r$, $q \equiv
\sqrt{(1-\mathcal{T}_B)/\mathcal{T}_B}$,
$s^{\gamma-1} q$  is the Fano parameter of the level $\gamma$.
Note that the lineshape $\mathcal{T}$ is reduced into the
single-level form~\eqref{FanoLineshape} when the level spacing
$\xi_2 - \xi_1 \ll \tilde{\Gamma}_\gamma$.
%i.e., when one of
%$\varepsilon_1$ and $\varepsilon_2$ is much larger than the other.

Figure~\ref{2NonintQD} shows typical two-level Fano
lineshapes as a function of
gate voltage $V_g$ for the cases where $U=0$ and the level broadening
$\tilde{\Gamma}_\gamma$ is comparable to the level spacing $\xi_2 -
\xi_1$. The entire lineshape may be understood as $s$-dependent
mixture of interferences between the paths through one QD level and
the direct channel (characterized by the Fano parameter of the
level) and between the paths through the two QD levels. For $s= -1$,
the upper resonance at $\xi_2$ has a negative value of Fano
parameter, while the lower one at $\xi_1$ has a positive value.
Therefore, the two resonances are out of phase (with difference by
$\pi$), as shown in Fig.~\ref{2NonintQD}(a). On the other hand, for
$s = +1$, the two resonances are in phase
[see Fig.~\ref{2NonintQD}(b)].

In Fig.~\ref{2NonintQD}, we plot the cross-correlation of level
occupation,~\cite{Konig05}
$\langle\delta n_1\delta n_2\rangle\equiv\langle
n_1n_2\rangle-\langle n_1\rangle\langle n_2\rangle\simeq-\langle
n_{12}\rangle^2$, which may give more
understanding of the $s$-dependent features.
For $s = +1$,
$\langle\delta n_1\delta n_2\rangle$
can have a finite value
and give rise to nonmonotonic behavior of $\langle n_\gamma \rangle$
[see $\langle n_1 \rangle$ around $eV_g = \xi_2$ in Fig.~\ref{2NonintQD}(b)],
while it vanishes for $s = -1$.
Note that it becomes suppressed
as $t_{LR}$ increases since the overlap between the two levels or
the level broadening $\tilde{\Gamma}_\gamma$ is reduced
[see Eq.~\eqref{Broadening}].

\begin{figure}[htb]
\includegraphics[width=0.4\textwidth]{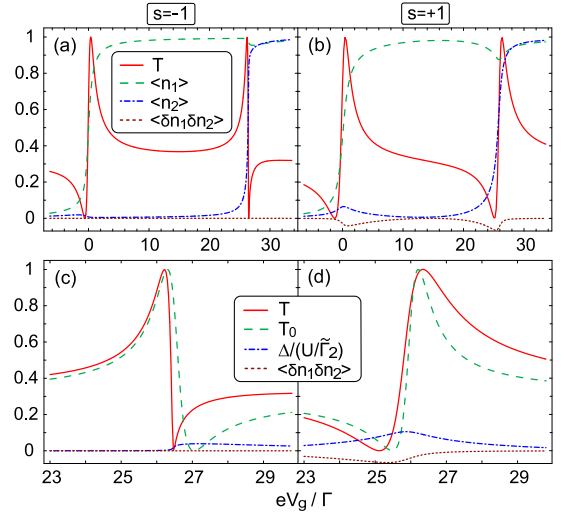}
\caption{(color online)
Upper panels: The same plots as in Fig.~\ref{2NonintQD}
in the interacting case of $U/\Gamma = 25$.
Lower panels:
The second resonances in (a) and (b) are analyzed
using the noninteracting Fano form $\mathcal{T}_0$ and
the nonmonotonicity $\Delta$
for (c) $s=-1$ and (d) $s=+1$, respectively (see text).
}\label{3IntQD}
\end{figure}

%\section{Fano lineshape modification by Coulomb repulsion}

Hereafter we turn on the Coulomb repulsion $U$ and discuss
how it modifies the resonance lineshape.
We find that within the SCHF treatment, the
lineshape has the same form as the noninteracting case of
$U=0$, Eq.~\eqref{NoninteractingFano},
\begin{equation}\label{InteractingFano}
\mathcal{T} = \mathcal{T}_B \frac{ [ \tilde{\varepsilon}_1
\tilde{\varepsilon}_2 - s - (\tilde{\zeta}^2-1) \delta_{s,1}
+ s q (\tilde{\varepsilon}_1 + s \tilde{\varepsilon}_2 - 2
\tilde{\zeta} \delta_{s,1} ) ]^2}{(
\tilde{\varepsilon}_1 \tilde{\varepsilon}_2 - s -
(\tilde{\zeta}^2 - 1) \delta_{s,1} )^2 + (\tilde{\varepsilon}_1
+ s \tilde{\varepsilon}_2 - 2 \tilde{\zeta} \delta_{s,1})^2},
\end{equation}
but with mean-field shifts
\begin{eqnarray}
\tilde{\varepsilon}_{\gamma}  =  \varepsilon_{\gamma} -
\frac{U}{\tilde{\Gamma}_{\gamma}}
\langle n_{\bar{\gamma}}\rangle, \, \, \, \, \,
\tilde{\zeta}  =  \zeta +
\frac{U}{\sqrt{\tilde{\Gamma}_1\tilde{\Gamma}_2}} \langle
n_{12}\rangle.
\label{Shift}
\end{eqnarray}
Note that $\varepsilon_\gamma$ and $\langle n_{\gamma \gamma'}
\rangle$ depend on $V_g$. The shift of the
detuning parameter can be understood as the level spacing
renormalization due to the Hartree repulsion. On the other
hand, the shift in $\tilde{\zeta}$ comes from the
Fock exchange, which is absent in the case of $s = -1$.

In Figs.~\ref{3IntQD}(a) and (b), we plot the lineshape $\mathcal{T}$
when nonmonotonic behavior~\cite{Berkovits05,Konig05,Sindel05,Goldstein06}
of $\langle n_{\gamma} \rangle$ occurs
(see, e.g., $\langle n_1 \rangle$ around the second resonance).
For $s=+1$, the nonmonotonic behavior comes from the Hartree
repulsion as well as from the Fock exchange,
while for $s=-1$, it is caused only by the former.
Therefore,
the $s=+1$ case shows the nonmonotonicity
in a wider range of $V_g$ where
$\langle \delta n_{1} \delta n_2 \rangle$ is enhanced by the Fock exchange.

The nonmonotonic dependence of $\langle n_{\gamma \gamma'} \rangle$
on $V_g$ modifies the lineshape
from the noninteracting cases. Such modification can be analyzed
when the level spacing renormalized by the Hartree contribution is
much larger than level broadening $\tilde{\Gamma}_\gamma$,
i.e., when $\xi_2 - \xi_1 + U \gg \tilde{\Gamma}_\gamma$,
so that the nonmonotonicity is not too strong. In this case, the
lineshape~\eqref{InteractingFano} can be simplified, for gate
voltage, for example, around the second resonance ($eV_g \simeq
\xi_2 + U  + s \pi\rho t_{LR}\tilde{\Gamma}_2$), into the
single-level Fano form $\mathcal{T}_0$,
\begin{eqnarray}\label{connection1}
\mathcal{T} (\tilde{\varepsilon}_2, q) \simeq
\mathcal{T}_0 (\tilde{\varepsilon}_2, q) \equiv
\mathcal{T}_B
\frac{(\tilde{\varepsilon}_{2}+ s q)^2}{\tilde{\varepsilon}_{2}^2+1},
\end{eqnarray}
but with not simple dependence of $\langle n_{1} \rangle$ on $V_g$
[see Eq.~\eqref{Shift}]. To analyze further,
we define nonmonotonicity measure
\begin{eqnarray}
\Delta (V_g) \equiv \frac{U}{\tilde{\Gamma}_2} ( \langle n_{1}
\rangle_{0} - \langle n_{1} \rangle) \simeq
\frac{U}{\tilde{\Gamma}_2} ( 1 -\langle n_{1} \rangle).
\end{eqnarray}
Here, $\langle n_{1} \rangle_{0} (V_g)$ is the level occupation
in the absence of the Coulomb repulsion, which can be obtained
from Eqs. (\ref{GDiagonal},\ref{GOffdiagonal},\ref{Occupation}).
We can take
$\langle n_{1}
\rangle_{0} (V_g) \simeq 1$ in this case of large renormalized
level spacing. For $| \Delta | \ll 1$, one has an approximated form
of the lineshape,
\begin{eqnarray}
\mathcal{T} & \simeq & \mathcal{T}_0 (\tilde{\varepsilon}_{2, 0}, q)
+ \frac{\partial \mathcal{T}_0 (\tilde{\varepsilon}_{2, 0}, q)}
{\partial \tilde{\varepsilon}_{2,0}} \Delta.
\label{connection2}
\end{eqnarray}
The leading term $\mathcal{T}_0$ obeys the single-level Fano
form~\eqref{FanoLineshape} with the detuning parameter
$\tilde{\varepsilon}_{2,0} = (eV_g-\xi_2-U\langle n_{1}
\rangle_{0})/\tilde{\Gamma}_2 + s \pi \rho t_{LR} \simeq
(eV_g-\xi_2-U) / \tilde{\Gamma}_2 + s \pi \rho t_{LR} $, which
approximately linearly depends on $V_g$, and the second deformation
term is {\em proportional} to $\Delta (V_g)$, therefore it provides
the connection between the lineshape and the nonmonotonicity.
Around the first resonance, one can find the same forms of
the connection and the nonmonotonicity measure but with
level index exchange
$1 \leftrightarrow 2$ and $\langle n_2 \rangle_{0} \simeq 0$.
The connection is applicable as well in the Breit-Wigner case
of $\mathcal{T}_{BW,0} = \mathcal{T}_0(t_{LR} \to 0)$.

In Figs.~\ref{3IntQD}(c) and (d), we plot the deviation of
$\mathcal{T}$ from $\mathcal{T}_0$ as well as $\Delta$.
The deviation depends on the phase parameter $s$, as the nonmonotonicity
does.
The lineshape deformation occurs in a wider range of $V_g$
in the case of $s= +1$, since
the Fock exchange enhances
$\langle \delta n_{1} \delta n_2 \rangle$ and therefore additional
nonmonotonicity only in the $s = +1$ case, as discussed before.
The connection between
the lineshape deformation and the nonmonotonicity of level
occupation found in Eq. (\ref{connection2}) may suggest an
experimental study of the nonmonotonic behavior.

\begin{figure}[t]
\includegraphics[width=0.35\textwidth]{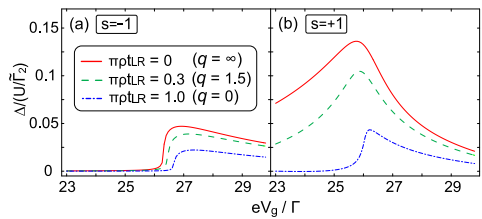}
\caption{(color online)
Nonmonotonicity $\Delta$ of level occupation as a function
of $V_g$
for (a) $s=-1$ and (b) $s=+1$
in the interacting case of $U/\Gamma = 25$.
Different values of $t_{LR}$ are chosen.
%the direct channel coupling $t_{LR}$ are chosen.
The other parameters are the same as in Fig. \ref{3IntQD}.
}
\label{4SideChannel}
\end{figure}

%\section{Discussion and summary}

We finally discuss a few remarks briefly.
First, the nonmonotonic behavior becomes weakened
for larger direct channel coupling $t_{LR}$, because
the level broadening $\tilde{\Gamma}_\gamma$ becomes narrower
[see Fig.~\ref{4SideChannel} and Eq.~\eqref{Broadening}].
Second, we extend the above findings to a spinful
single-level QD.
In this case, the Hartree repulsion induces the
nonmonotonic behavior of level occupation,
but there is no interlevel interference and no Fock contribution.
We find
that the connection~\eqref{connection2} between the lineshape
deformation and the nonmonotonicity is still hold
(at temperatures larger than the Kondo temperature).

Third, the temperature range where the SCHF approach is valid
depends on $t_{LR}$.
It has been found~\cite{Lee07,Kashcheyevs07,Silvestrov07} that
when $t_{LR} = 0$, the two-level QD can be mapped into a Kondo
system and it can show correlation-induced resonance~\cite{Meden06}
below the Kondo temperature. We find that the two-level QD with the
direct channel can be also mapped~\cite{Lee}
into a Kondo system, depending on $s$,
and that the corresponding Kondo temperature
decreases with increasing $t_{LR}$.
Therefore, our approach may be valid above the Kondo
temperature, the upper bound of which can be estimated~\cite{Lee07}
from the case of $t_{LR} = 0$.

%In summary, we have studied the two-level Fano lineshape in the
%presence of Coulomb repulsion and analyzed
%Coulomb modification of the lineshape.

In summary, we have studied Fano resonance lineshape and the
nonmonotonicity of level occupation in a two-level QD side-coupled
two leads. The two-level lineshape is derived for both the
noninteracting and interacting cases [Eqs.
(\ref{NoninteractingFano},\ref{InteractingFano})]. We especially
obtain the connection, Eq. (\ref{connection2}), between the
nonmonotonicity and the Coulomb modification of the lineshape. We
also find that stronger direct-channel coupling weakens the
nonmonotonicity.

This work was supported by a Korean Research Foundation Grant
(KRF-2006-331-C00118).

\end{document}